\documentclass[sigconf,nonacm, authorversion]{acmart}

\AtBeginDocument{%
  }
    
\setcopyright{acmlicensed}
\copyrightyear{2025}
\acmYear{2025}
\acmDOI{XXXXXXX.XXXXXXX}

\acmConference[EASE 2025]{The 29th International Conference on Evaluation and Assessment in Software Engineering}{17–20 June, 2025}{Istanbul, Türkiye}

\begin{document}
\title{Towards Lean Research Inception: Assessing Practical Relevance of Formulated Research Problems}

\author{Anrafel Fernandes Pereira}
\affiliation{%
  \institution{PUC-Rio}
  \city{Rio de Janeiro}
  \country{Brazil}}
\email{afpereira@inf.puc-rio.br}

\author{Marcos Kalinowski}
\affiliation{%
  \institution{PUC-Rio}
  \city{Rio de Janeiro}
  \country{Brazil}}
\email{kalinowski@inf.puc-rio.br}

\author{Maria Teresa Baldassarre}
\affiliation{%
  \institution{University of Bari}
  \city{Bari}
  \country{Italy}}
\email{mariateresa.baldassarre@uniba.it}

\author{Jürgen Börstler}
\affiliation{%
  \institution{Blekinge Institute of Technology}
  \city{Karlskrona}
  \country{Sweden}}
\email{jurgen.borstler@bth.se}

\author{Nauman bin Ali}
 \affiliation{%
   \institution{Blekinge Institute of Technology}
   \city{Karlskrona}
   \country{Sweden}}
 \email{nauman.ali@bth.se}

\author{Daniel Mendez}
\affiliation{%
  \institution{Blekinge Institute of Technology}
  \city{Karlskrona}
  \country{Sweden}}
\email{daniel.mendez@bth.se}

\renewcommand{\shortauthors}{Pereira \textit{et al.}}

\begin{abstract}
[Context] The lack of practical relevance in many Software Engineering (SE) research contributions is often rooted in oversimplified views of industrial practice, weak industry connections, and poorly defined research problems. Clear criteria for evaluating SE research problems can help align their value, feasibility, and applicability with industrial needs. [Goal] In this paper, we introduce the Lean Research Inception (LRI) framework, designed to support the formulation and assessment of practically relevant research problems in SE. We describe its initial evaluation strategy conducted in a workshop with a network of SE researchers experienced in industry-academia collaboration and report the evaluation of its three assessment criteria (\emph{valuable}, \emph{feasible}, and \emph{applicable}) regarding their importance and completeness in assessing practical relevance. [Method] We applied LRI retroactively to a published research paper, engaging workshop participants in discussing and assessing the research problem by applying the proposed criteria using a semantic differential scale. Participants provided feedback on the criteria's importance and completeness, drawn from their own experiences in industry-academia collaboration. [Results] The findings reveal an overall agreement on the importance of the three criteria~--~\textit{valuable} (83.3\%), \textit{feasible} (76.2\%), and \textit{applicable} (73.8\%)~--~for aligning research problems with industrial needs. Qualitative feedback suggested adjustments in terminology with a clearer distinction between \textit{feasible} and \textit{applicable}, and refinements for \textit{valuable} by more clearly considering \textit{business value}, \textit{ROI}, and \textit{originality}. [Conclusion] While LRI still constitutes ongoing research and requires further evaluation, our emerging results strengthen our confidence that the three criteria applied using the semantic differential scale can already help the community assess the practical relevance of SE research problems.
\end{abstract}



\maketitle

\section{Introduction}
The practical relevance of Software Engineering (SE) research is often limited by a disconnect between academia and industry, with studies failing to address practically relevant challenges despite their theoretical contributions~\cite{garousi2020, franch2020, winters2024}. Gorschek \textit{et al.}~\cite{gorschek2006, gorschek2021solving} emphasize that a well-defined problem formulation is critical for research to generate industrial impact. Many research problems are formulated in isolation, lacking input from industrial practice~\cite{gorschek2021solving}. This disconnect hinders understanding of real and practical industry challenges. Ensuring practical relevance requires structured approaches that align research with industry needs through clear problem formulation and assessment~\cite{ali2016, molleri2023, storey2024disruptive}.

To integrate academic and industry perspectives early in the research process, we introduce the Lean Research Inception (LRI) framework to support the formulation and initial assessment of practically relevant SE research problems. In this paper, we present the complete vision of the LRI framework and an evaluation of its research problem assessment phase. The assessment employs a semantic differential scale~\cite{heise1970} to assess the relevance of formulated research problems based on three criteria: \textit{valuable} (whether solving the problem can generate meaningful impact for industrial practice), \textit{applicable} (whether the problem leads to practical and usable outcomes in industry), and \textit{feasible} (whether the problem can realistically be investigated given available resources). The scale was inspired by agile principles and Lean Startup's Minimum Viable Product (MVP) concept~\cite{ries2011}, aiming to enhance the effectiveness of aligning research with industry needs.

The empirical evaluation was conducted through a workshop with 42 senior SE researchers at the 2024 annual meeting of the International Software Engineering Research Network (ISERN\footnote{https://isern.iese.de/}). The study focused on evaluating the three key criteria~--~\textit{valuable, feasible, and applicable}~--~used in the semantic differential scale. Participants analyzed these criteria by discussing an example research problem, capturing their insights through collaborative annotations, and individually applying the criteria to assess the problem's relevance. Following this discussion, they completed a survey on the perceived importance and completeness of each criterion. The survey provided quantitative and qualitative insights into how well these criteria are perceived as important to support an initial assessment of the practical relevance of formulated research problems.

The results indicate that our criteria, \textit{valuable} (83.3\%), \textit{feasible} (76.2\%), and \textit{applicable} (73.8\%) are appropriate for assessing research problem relevance, with high levels of agreement among participants. Furthermore, qualitative feedback suggested adjustments in terminology for a clearer distinction between \textit{feasible} and \textit{applicable}, and refinements for \textit{valuable} to include concrete examples to ease understanding, such as \textit{business value}, \textit{ROI}, and \textit{originality}.

\section{Related Work}
Software Engineering research is primarily aimed at industrial applications, which makes alignment with industry needs crucial for practical relevance \cite{winters2024}. Achieving this requires strategies that balance theoretical advancements with real-world applicability \cite{stol2018}. Ivarsson \textit{et al.}~\cite{ivarsson2011} proposed a model to assess the rigor and industrial relevance of technology evaluations in SE. In their analyses, the model revealed that most evaluations lacked both rigor and industrial relevance. Additionally, the study found no significant improvement in industrial relevance over time.

Garousi \textit{et al.}~\cite{garousi2020} investigate the practical relevance of SE research, synthesizing community opinions and evidence through a multivocal literature review (MLR). They identify three key factors limiting relevance: too simplistic views of practice, weak industry connections, and poor problem identification. To address this, they recommend adopting appropriate research approaches, focusing on practical problems, and strengthening industry collaboration. The study emphasizes the need for rigorous empirical research to better align SE studies with industrial needs. Winters~\cite{winters2024} further highlights the misalignment between SE research and industry needs, criticizing the lack of practical context, narrow problem scopes, and limited scalability, emphasizing the need for research that delivers measurable and practical value to industry.

Petersen \textit{et al.}~\cite{petersen2024} propose a reasoning framework to enhance the design, assessment, and reporting of industrial relevance in SE research. Given the lack of consensus on defining and measuring relevance, they review key attributes such as applicability, context, and practical impact. Their framework, structured around six aspects \textit{(what, how, where, who, when, and why)}, provides an approach for evaluating and communicating research relevance of applied research in industry contexts. Storey \textit{et al.}~\cite{storey2024disruptive} propose a playbook for researching disruptive innovations in SE. They emphasize interdisciplinary collaboration, industry engagement, and iterative experimentation while underscoring the need for clear criteria to assess impact and applicability in industrial and social contexts.

These studies emphasize the importance of aligning SE research with industry needs, reinforcing the motivations behind our work. Ivarsson \textit{et al.}~\cite{ivarsson2011} highlight the persistent lack of rigor and industrial relevance in SE evaluations, underscoring the need for structured approaches to bridge this gap. Garousi \textit{et al.}~\cite{garousi2020} and Winters~\cite{winters2024} critique the disconnect between academia and industry, pointing to weak collaboration and impractical research focus—challenges LRI directly addresses by integrating practitioners early in the research process. Petersen \textit{et al.}~\cite{petersen2024} propose a reasoning framework for industrial relevance, which helped to define LRI's problem formulation attributes and assessment criteria. Finally, Storey \textit{et al.}\cite{storey2024disruptive} provide valuable insights into assessing the impact of disruptive innovations. 

LRI advances previous work by proposing an iterative framework that integrates academic and industry perspectives from the outset to support the formulation of research problems and an initial assessment of their practical relevance through clear and easy-to-apply criteria.

\section{Lean Research Inception}
The Lean Research Inception (LRI) framework was developed to help bridging the gap between SE research and industrial practice. Grounded in agile principles and methodologies such as Design Thinking~\cite{plattner2009}, Lean Startup~\cite{ries2011}, and Lean Inception~\cite{caroli2017}, LRI fosters collaboration between researchers and practitioners from the very beginning of a research project. Its core objective is to integrate these stakeholders early on to ensure that research problems are practically relevant and applicable to real-world scenarios.

LRI complements the technology transfer model proposed by Gorschek \textit{et al.}~\cite{gorschek2006}, which defines seven stages for transitioning research into industrial practice. LRI focuses on how to engage in one of the most critical stages: ``Problem Formulation''. LRI refines this stage with a structured and transparent process. It includes practical recommendations~\cite{garousi2016challenges} to promote early alignment between academia and industry. The approach actively involves SE researchers and practitioners from the beginning. However, it acknowledges that engaging practitioners is challenging. By promoting this collaborative approach, LRI aims to enhance the relevance and applicability of SE research, fostering a more effective exchange between academia and industry.

\begin{figure}
    \centering
    \includegraphics[width=1\linewidth]{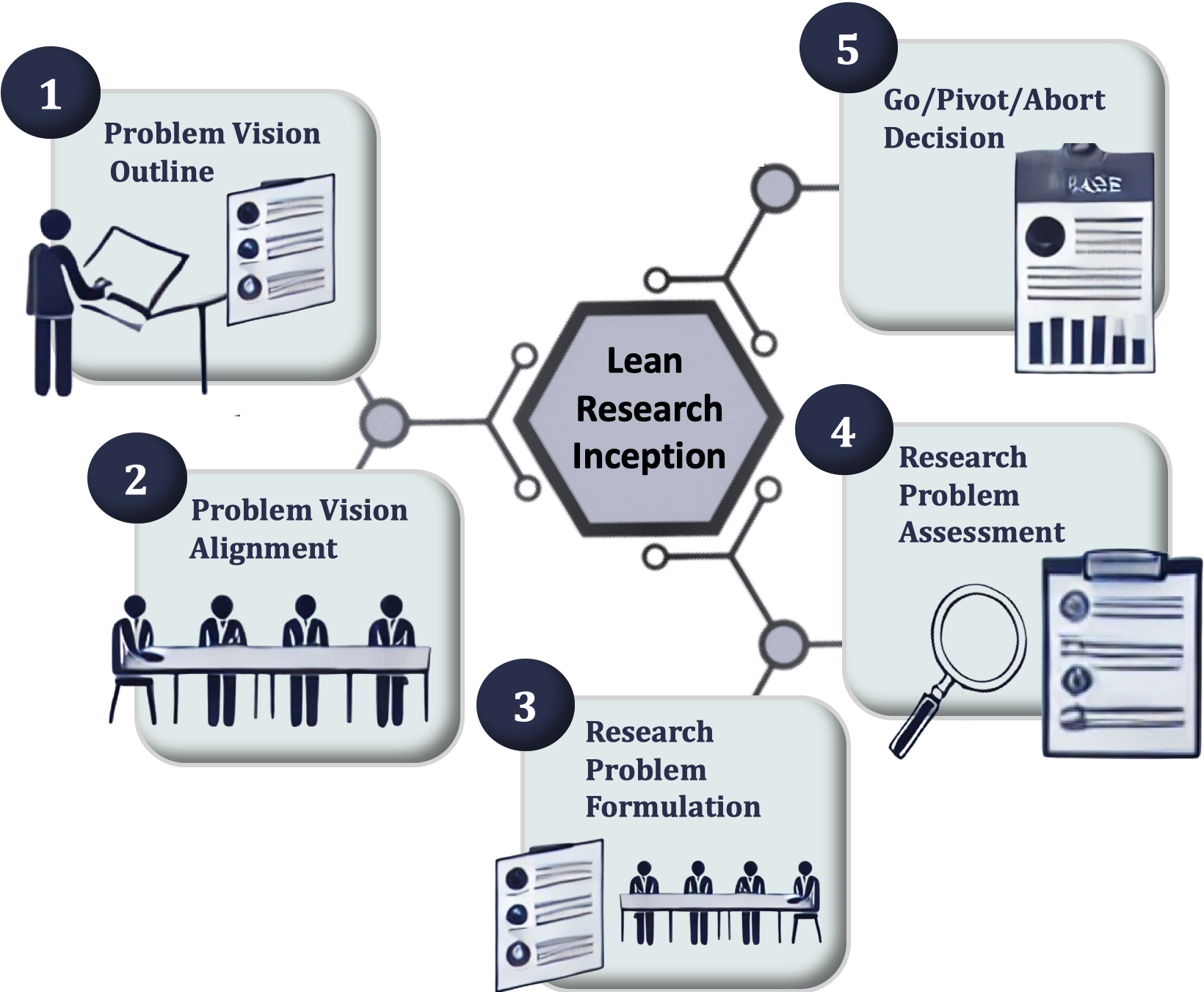}
    \caption{Lean Research Inception Overview}
    \label{fig:lri_overview}
\end{figure}

The LRI framework comprises five sequential phases, as illustrated in Figure \ref{fig:lri_overview}. We created an interactive board template that can be used to guide these phases and be filled out collaboratively in a shared workspace. Our open science repository~\cite{zenodoRepository} provides a blank template that allows visualizing the framework in detail and a complete example filled out based on a published research paper.

\textbf{Phase 1 - Problem Vision Outline:} SE researchers work collaboratively to create an initial draft of the practical problem by filling attributes of the ``Problem Vision'' board (\textit{problem outline} (what/how/why), \textit{context} (where/when), \textit{implications/impacts} (why), \textit{practitioners} (who), \textit{evidence} (how), \textit{objective} (what/how), and \textit{research questions} (what)). The attributes were selected based on previous research on relevance in SE, also considering their support in the broader literature on SE, especially regarding their role in problem formulation~\cite{garousi2020, gorschek2006, ivarsson2011, petersen2024}.
    
\textbf{Phase 2 - Problem Vision Alignment:} Practitioners join the SE researchers in a collaborative workshop. After introductions, the researchers present the initial ``Problem Vision''. The participants then review, discuss, and refine its seven attributes to ensure a shared understanding and improve the problem formulation.

\textbf{Phase 3 - Research Problem Formulation:} Researchers and practitioners refine and document the research problem based on the discussions in Phase 2. The formulated research problem serves as the basis for the assessment in the next phase.

\textbf{Phase 4 - Research Problem Assessment:} Participants individually assess the relevance of the formulated research problem using a semantic differential scale~\cite{heise1970}. Adapted to LRI, this scale assesses three key criteria of practical relevance in SE, aligned with agile principles and the concept of a Minimum Viable Product (MVP)~\cite{ries2011}. They are: \textit{``Worthless - Valuable''}, which assesses whether solving the problem can generate meaningful impact for industrial practice; \textit{``Infeasible - Feasible''}, which assesses whether the problem can realistically be investigated given available resources; and \textit{``Inapplicable - Applicable''}, which assesses if the problem leads to practical and usable outcomes for the industry. Each criterion is rated on a seven-point scale, enabling a detailed analysis.

\textbf{Phase 5 - Go/Pivot/Abort Decision:} In this final phase, the group assessment results are consolidated. Aligned with the MVP-approach~\cite{ries2011}, LRI seeks to ensure that the research problem is relevant (\textit{valuable}, \textit{feasible}, and \textit{applicable}) to real-world SE scenarios. Similarly to MVPs, if the perceived relevance is high, the investigation should continue (\textit{e.g.}, to the `Candidate Solution' stage of the Gorschek \textit{et al.}'s model~\cite{gorschek2006}). If the perceived relevance is low but adjustable, the research problem should be realigned (or pivoted), restarting at Phase 2. However, if the perceived relevance is low and cannot be fixed, the investigation should be aborted to prevent wasted effort on irrelevant research.

\section{Evaluation}\label{sec:evaluation}
We conducted an empirical study with 42 senior SE researchers experienced in industry-academia collaboration to evaluate three criteria—\textit{valuable}, \textit{feasible}, and \textit{applicable}—in assessing the practical relevance of research problems. As part of our initial evaluation strategy, we enacted the LRI framework within the ISERN workshop. Participants were introduced to a pre-filled ``Problem Vision'' board, based on a published research problem~\cite{cabral2024}, to explore its structure and attributes (Phase 1). They then discussed and refined the problem collaboratively (Phases 2 and 3), followed by an individual assessment using the semantic differential scale (Phase 4). The exercise concluded with the consolidation of perceptions and submission of individual feedback via a survey (Phase 5).

This study offers a complete view of the LRI framework, with a specific focus on Phase 4, which evaluates the practical relevance of formulated research problems. The analysis is based on data collected from survey responses on the perceived importance and completeness of the three evaluation criteria. While this represents a small-scale evaluation rather than a full case study~\cite{wohlin2022case}, this approach tends to be well-suited for the initial assessment of the prospective value of a new proposal~\cite{wohlin2022case}. We adhered to the case study reporting structure recommended by Runeson \textit{et al.}\cite{runeson2012case}.

\subsection{Goal and Research Questions}
We define the goal of this study using the goal definition template of the Goal Question Metric (GQM) paradigm~\cite{basili1988} as follows:

``\textbf{Analyze} the criteria (\textit{valuable}, \textit{feasible}, and \textit{applicable}) of the semantic differential scale \textbf{for the purpose of} characterization \textbf{with respect to} the perception of their importance and completeness to support practical relevance assessment of formulated research problems \textbf{from the point of view of} senior SE researchers with experience in industry-academia collaborations \textbf{in the context of} conducting a practical relevance assessment and providing feedback on the importance of the criteria.''

Therefrom, we formulate the following two research questions:
\begin{itemize}
    \item \textbf{RQ1:} To what degree are the three criteria perceived as important to support practical relevance assessment of formulated research problems?
    
    \item \textbf{RQ2:} What other criteria are perceived as important to support practical relevance assessment of formulated research problems?
\end{itemize}

\subsection{Case and Subject Selection}\label{sec:participants}
We conducted this study during an in-person workshop at ISERN 2024 in Barcelona, Spain, engaging 42 senior SE researchers with experience in industry-academia collaboration. The setting provided a valuable opportunity to explore our objectives by leveraging participants’ expertise in a collaborative environment. We retrospectively applied the LRI phases using a research problem from a published article~\cite{cabral2024}. We selected the research problem from this paper as the discussion case because it explores the application of familiar software design principles to AI-enabled systems and was recognized with an award for its practical relevance~\cite{staron2024bringing}.

\subsection{Instrumentation}\label{sec:artifacts}
We carefully designed and reviewed all materials for this study to ensure the reliability of data collection. To support this process, we conducted a pilot workshop with twelve master’s and Ph.D. students experienced in Experimental and Empirical SE at ExACTa\footnote{https://www.exacta.inf.puc-rio.br} PUC-Rio. Their familiarity with empirical research enabled them to provide valuable feedback on the clarity, consistency, and usability of the artifacts before their use in the main study.

The study instrumentation consisted of a slide presentation and three main artifacts structured in three envelopes each corresponding to a specific activity, to guide participants through a step-by-step collaborative process. The first envelope included a pre-filled A2-format ``Problem Vision'' board and the article by Cabral \textit{et al.}~\cite{cabral2024}, which participants reviewed and discussed in the first activity, documenting their insights on the board using post-its. The second envelope included a semantic differential scale~\cite{heise1970} to assess the previously discussed research problem based on value, applicability, and feasibility, which participants applied during the second activity. The third envelope included a Likert-type scale survey with optional open-ended questions to evaluate the importance and completeness of LRI's attributes and criteria, and to gather suggestions for additional elements not covered by the framework. Participants were divided into five heterogeneous groups to ensure diverse perspectives and balanced discussions. Each group received the three envelopes. All materials are anonymously available in our open science repository~\cite{zenodoRepository}.

\subsection{Data Collection and Analysis Procedures}
We collected both quantitative and qualitative feedback using the outlined artifacts. Our mixed-methods approach combined statistical analysis of Likert-type scale responses with qualitative analysis of open-ended feedback to comprehensively assess participant perceptions. For the quantitative analysis, we calculated agreement level frequencies for each criterion. For the qualitative analysis, we examined open-ended responses to support the quantitative findings, capture nuances, and identify divergences. All data was recorded in a spreadsheet, available in our open science repository~\cite{zenodoRepository}.

\section{Results}
\subsection{Case and Subject Description}
The evaluation was carried out through a 90-minute collaborative workshop at ISERN 2024, structured into five steps: (i) an introductory presentation (20 min); (ii) group formation and material distribution (5 min); (iii) discussion of the research problem using the ``Problem Vision'' board (35 min); (iv) assessment of the relevance of the formulated problem using the semantic differential scale (15 min); and (v) a survey (15 min).

\subsection{Perceived importance of criteria (RQ1)}
\begin{figure}
    \centering
    \includegraphics[width=1\linewidth]{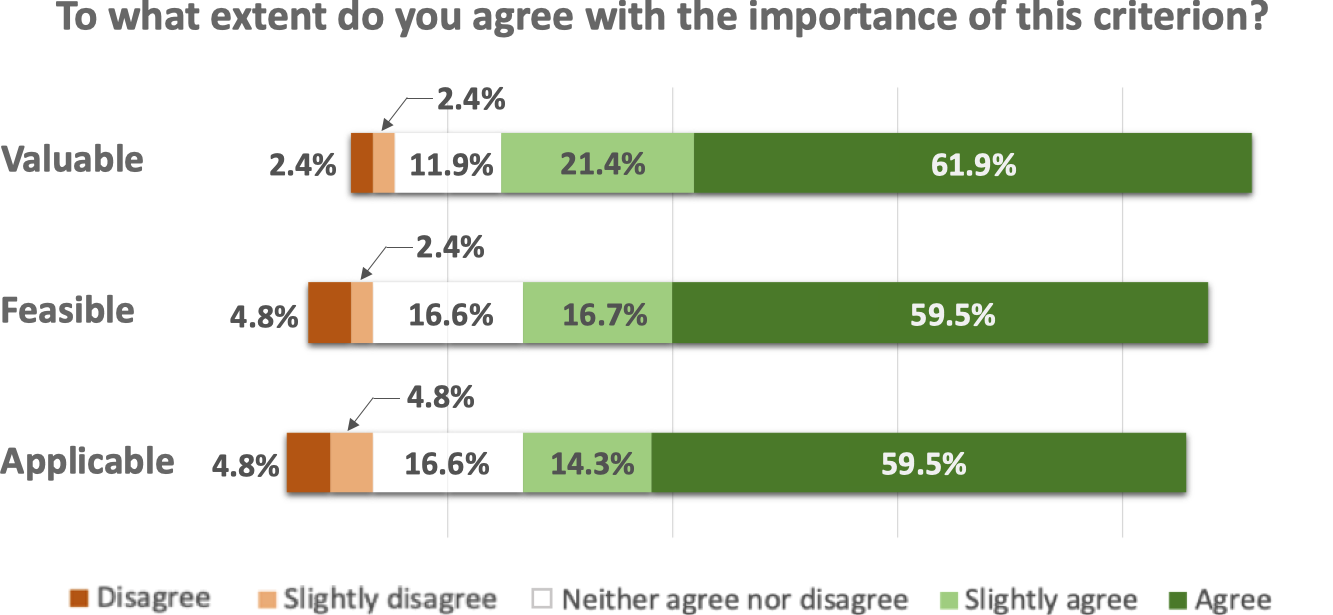}
    \caption{Questionnaire results}
    \label{fig:chart}
\end{figure}

The quantitative analysis of the collected data, shown in Figure~\ref{fig:chart}, revealed a general trend indicating the importance of the three criteria as dimensions of the semantic differential scale. In the following, we present a detailed criterion-by-criterion analysis.

\textbf{Valuable:} The results for the \textit{valuable} criterion demonstrated acceptance, with 83.3\% of participants agreeing to some extent with its importance (21.4\% Slightly Agree and 61.9\% Agree). Only 4.8\% expressed some level of disagreement (Disagree and Slightly Disagree), while 11.9\% remained neutral (Neither Agree nor Disagree). Qualitative feedback reinforces this perception. Participants emphasized that research should focus on meaningful and practical problems. P15 noted, \textit{``Research can be done that no one cares about; the research should be conducted with a salient problem of stakeholders,''} highlighting the importance of addressing real-world issues. Similarly, P35 stated, \textit{``No point in doing something with no utility,''} reinforcing the need for research to provide tangible benefits.

Despite this consensus, some participants suggested refinements. P11 recommended reconsidering the term \textit{``Worthless''} in the scale, stating, \textit{``Lower end of the scale `Worthless' sounds too negative. Consider an alternative word?''} P16 suggested a clearer definition: \textit{``I suggest providing a definition of the criteria,''} while P27 proposed breaking it down into specific elements: \textit{``I would suggest breaking down the criterion into elements that can be used to make a useful decision.''} Further concerns about subjectivity emerged among neutral and disagreeing participants. P28 described the criterion as \textit{``very subjective,''} while P20 found it \textit{``too fuzzy,''}, indicating a need for more precise wording and refinements to improve clarity.

\textbf{Feasible:} The \textit{feasible} criterion also received mainly positive results, with 76.2\% of participants agreeing to some extent with its importance (16.7\% Slightly Agree and 59.5\% Agree). Meanwhile, 16.6\% remained neutral, and 7.2\% expressed some level of disagreement (2.4\% Slightly Disagree and 4.8\% Disagree). Qualitative feedback highlighted the importance of feasibility for applied research. P25 stated, \textit{``Key for applied research,''} and P15 emphasized feasibility’s influence on early investigative stages: \textit{``The feasibility can influence the `investigation' step incrementally.''}Nonetheless, some participants reported difficulties in fully assessing the criterion. For example, P37 noted, \textit{``It is important but I missed information to answer it,''}. P40 questioned, \textit{``From research or industry perspective?''} suggesting that the research perspective should be clearer stated. Those who were neutral or slightly disagreed expressed difficulties in applying the criterion. P17 mentioned the need for clarifying available resources: \textit{``Clarify available resources.''} P39 raised difficulties in assessing feasibility during problem definition: \textit{``Is it possible to assess this dimension at the moment of problem definition?''}

\textbf{Applicable:} The results for \textit{applicable} almost mirrored those of \textit{feasible}. Approximately 73.8\% of participants agreed with its importance (14.3\% Slightly Agree and 59.5\% Agree), while 16.6\% were neutral, and 9.6\% expressed some level of disagreement (4.8\% Slightly Disagree and 4.8\% Disagree). Participants who agreed emphasized its importance while noting limitations in its assessment. P24 commented, \textit{``Important, but the scale is not helpful,''} while P06 observed that feedback provided for other criteria also applied to \textit{Applicable}. Neutral participants raised concerns about its clarity and practicality. For example, P35 stated, \textit{``This maps to it being useful,''} while P39 questioned, \textit{``I wonder how easy it is to assess this dimension at the moment of defining the problem.''} Disagreeing participants raised foundational issues with the criterion. P20 described it as \textit{``Too fuzzy,''} and P13 noted that applicability could only be determined after stakeholders are convinced of a problem’s relevance: \textit{``Applicability is independent of relevance. Deciding applicability only be possible once a company was convinced of the problem’s relevance.''} Additionally, P31 observed a strong correlation between \textit{feasible} and \textit{applicable}: \textit{``They are very correlated. If something is unfeasible, it would be directly inapplicable.''}. This indicates an opportunity to provide a clearer distinction between these two criteria.

Overall, the quantitative and qualitative findings confirm that all three criteria—\textit{valuable, feasible, and applicable}—are mainly perceived as important for the assessment of research problem relevance. However, they also highlight opportunities for refinement, particularly in improving the clarity and contextualization of the criteria to enhance their effectiveness in future evaluations.

\subsection{Completeness of criteria (RQ2)}
To answer RQ2, participants suggested additional criteria they consider important for assessing practical relevance. The following analysis summarizes their input.

\textbf{Business value and technical impact:} Participants P19 and P20 emphasized the importance of distinguishing between \textit{business value} and \textit{technical impact}. While aligned with the \textit{valuable} criterion, their suggestions highlight the need to evaluate economic and strategic benefits separately from technical contributions.

\textbf{Measurable benefits:} P14 proposed including \textit{``expected measurable benefit,''} which resonates with the \textit{valuable} criterion. This suggestion adds depth by emphasizing the need for clear, quantifiable outcomes, potentially improving the precision with which value is assessed.

\textbf{Originality:} P40 suggested assessing \textit{originality} to ensure research problems explore new areas. Although not explicitly covered in the current criteria, \textit{originality} complements \textit{valuable} by emphasizing innovation, helping to distinguish impactful problems from those that replicate existing solutions.

\textbf{Cost and Return on Investment (ROI):} P24 and P35 highlighted \textit{cost} and \textit{ROI} as essential factors. This economic perspective partially aligns with \textit{feasible} and partially with \textit{valuable}, emphasizing the need to consider resource efficiency explicitly.

\textbf{Impact and limitations:} P23 and P37 emphasized the importance of assessing both the \textit{impact} and \textit{limitations} of solutions. While aligning with \textit{valuable} and \textit{feasible}, these factors help balance potential benefits with implementation challenges. P37 specifically highlighted the need to clarify limitations to better assess feasibility, stating, \textit{``I think that main limitations/restrictions should be made clear in order to reflect on feasibility.''}

\textbf{Applicability requirements:} P28 proposed defining \textit{``Requirements for the applicability''}, suggesting defining specific thresholds to make \textit{applicable} more measurable.

\textbf{Problem importance:} P07 questioned: \textit{``Is this a problem worth putting resources into?''}. This consideration aligns with \textit{valuable} and \textit{feasible}, balancing problem importance with the resources needed for its resolution.

\textbf{Intention to use:} P25 suggested assessing intention to use, based on the Technology Acceptance Model. This recommendation further details \textit{applicable} by emphasizing end-user adoption.

Overall, the additional recommendations overlap or refine existing criteria.\textit{Valuable} may consider \textit{business value}, \textit{measurable benefits}, \textit{originality}, \textit{ROI}, and \textit{problem importance}; \textit{feasible} involves analyzing \textit{costs} and \textit{limitations}; and \textit{applicable} gains specificity through \textit{applicability requirements} and \textit{intention to use}. 

\section{Discussion}
Quantitative results showed high levels of agreement (slightly agree and agree) on the importance of the criteria, especially for \textit{valuable} (83.3\%), followed by \textit{feasible} (76.2\%) and \textit{applicable} (73.8\%). Hence, our initial results highlight the importance of the semantic differential scale criteria (\textit{valuable, feasible, applicable}) for an initial assessment of research relevance. Qualitative feedback refined these findings, suggesting improvements in terminology definitions. For instance, the participants suggested a clearer distinction between \textit{feasible} and \textit{applicable}. This could be solved by better explaining that feasibility takes the research perspective, while applicability considers the industry perspective.

Participants suggested refinements rather than entirely new criteria, focusing on aspects of \textit{valuable} (\textit{e.g.}, \textit{business value}, \textit{measurable benefits}, \textit{originality}, \textit{ROI}, \textit{problem importance}), \textit{feasible} (\textit{e.g.}, \textit{costs}, \textit{limitations}), and \textit{applicable} (\textit{e.g.}, \textit{applicability requirements}, \textit{intention to use}). These suggestions reflect a need to clarify value, add contextual detail for feasibility, and define measurable applicability criteria. However, as noted by some participants, evaluating these dimensions precisely during problem formulation is difficult. Thus, we consider the MVP-inspired assessment scale suitable for early-stage evaluation, enabling flexible, collective judgment before more detailed metrics are applied.

The contribution of this work complements related work focusing on assessing the practical relevance of research problems in SE. While Garousi \textit{et al.}~\cite{garousi2020} highlight gaps in practical relevance and Winters~\cite{winters2024} criticizes the limited applicability of academic studies, LRI addresses these concerns by integrating practitioners early in the process of problem formulation and relevance assessment. In line with the need of assessing relevance described by Ivarsson \textit{et al.}~\cite{ivarsson2011}, and differently from the more complete reasoning framework to assess industrial relevance~\cite{petersen2024}, inspired by the desired properties of MVP \cite{ries2011}, LRI defines a simple semantic differential scale with three main criteria~–\textit{valuable, feasible and applicable}–~that can be used to support an initial collective assessment of the relevance of formulated research problems. 

\section{Threats to Validity}
We addressed the four categories of validity threats described by Wohlin \textit{et al.}~\cite{wohlin2024}:

\textbf{Internal Validity:} To mitigate threats to internal validity, we randomly assigned participants to groups, and each participant individually applied the assessment scale for the selected case before answering the survey questions on the assessment criteria. Furthermore, the survey provided open-ended questions for justification and allowed participants to suggest additional criteria.

\textbf{External Validity:} Conducting the study with a group of participants that may limit generalizability beyond (senior) SE researchers. However, we argue that ISERN members, as per ISERN's mission, are expected to be experts in empirical SE and industry-academia collaboration, and thus render this group highly relevant for this study. Hence, we consider these emerging results to be useful initial indications. Nevertheless, future research should involve industry stakeholders for broader generalizability.

\textbf{Construct Validity:} The Likert-type scale may not have fully captured perceptions, and interpretations of the criteria could vary. We addressed this by including open-ended questions, carefully reviewing workshop materials, and conducting a pilot study with master and Ph.D. students to refine study instruments.

\textbf{Conclusion Validity:} The sample size (42 participants) and reliance on Likert-type scale responses limit inferential statistics. To strengthen the findings, we triangulated quantitative data with qualitative analysis, ensuring a more comprehensive interpretation.

\section{Concluding Remarks}
This study introduced the Lean Research Inception (LRI) framework as a structured approach to support the formulation and initial assessment of practically relevant research problems in SE. We evaluated the three criteria of LRI’s Semantic Differential Scale in terms of importance and completeness to support this initial assessment. Our findings indicate an overall agreement on the importance of the three criteria~--~\textit{valuable} (83.3\%), \textit{feasible} (76.2\%), and \textit{applicable} (73.8\%)~--~in aligning research problems with industrial needs. Qualitative feedback suggested refining terminology to better distinguish between \textit{feasible} and \textit{applicable} and expanding \textit{valuable} to encompass \textit{business value}, \textit{ROI}, and \textit{originality}.

Although LRI remains an ongoing research effort requiring further evaluation, these emerging results indicate that early practitioner involvement and the use of the semantic differential scale offer a practical approach for the initial assessment of the relevance of SE research problems. Future research could refine the terminology of the scale and include practical examples for each dimension. In addition, more industry-driven case studies are needed to evaluate the effectiveness of LRI in real-world collaboration contexts.

\begin{acks}
We thank the 42 SE researchers from ISERN who voluntarily participated in our study. This work has been partially supported by the following projects: SERICS, grant PE00000014, MUR National Recovery and Resilience Plan, European Union - NextGenerationEU (EU-NGEU); QUASAR, grant 2022T2E39C, PRIN 2022 MUR program, EU-NGEU; S.E.R.T. Research Profile project, KKS foundation; CNPq, grant 312275/2023-4; FAPERJ, grant E-26/204.256/2024; and CAPES, finance code 001.
\end{acks}

\bibliographystyle{ACM-Reference-Format}
\bibliography{references}

\appendix
\end{document}